\documentclass[aps,twocolumn,superscriptaddress]{revtex4}
\input epsf

\begin{document}

\title{Bulk motion of granular matter in an agitated cylindrical bed}

\author{Phanikumar Sistla}
\affiliation{Department of Mechanical and Materials Engineering, 
University of Western Ontario, London, Ontario N6A~5B9, Canada}

\author{Oleh Baran}
\affiliation{Department of Applied Mathematics,
University of Western Ontario, London, Ontario N6A~5B7, Canada}

\author{Q. Chen}
\affiliation{Department of Mechanical and Materials Engineering, 
University of Western Ontario, London, Ontario N6A~5B9, Canada}

\author{Peter H. Poole} 
\affiliation{Department of Applied Mathematics, University of Western
Ontario, London, Ontario N6A~5B7, Canada}
\affiliation{Department of Physics, St. Francis Xavier University,
Antigonish, Nova Scotia B2G~2W5, Canada}

\author{Robert J. Martinuzzi}
\affiliation{Department of Mechanical and Materials Engineering, 
University of Western Ontario, London, Ontario N6A~5B9, Canada}

\date{\today}

\begin{abstract}
Experimental results are reported for the bulk motion induced in a bed
of granular matter contained in a cylindrical pan with a flat bottom
subjected to simultaneous vertical and horizontal vibrations.  The
motion in space of the moving pan is quantified.  A number of distinct
bulk dynamical modes are observed in which the particle bed adopts
different shapes and motions. At the lowest pan excitation frequency
$\omega$, the bed forms a ``heap,'' and rotates about the cylinder
axis.  As $\omega$ is increased, a more complex ``toroidal'' mode
appears in which the bed takes the shape of a torus; in this mode,
circulation occurs both about the cylinder axis, and also radially,
with particles moving from the outer edge of the pan to the centre on
the top surface of the bed, and back to the outer edge along the pan
bottom.  At the highest $\omega$, surface modulations (``surface
waves'' and ``sectors'') of the toroidal mode occur.  The origin of
this family of behavior in terms of the pan motion is discussed.
\end{abstract}

\maketitle

\section{INTRODUCTION}

Granular materials play an important role in many industrial
processes. In applications, granular materials are often subjected to
vibrations to generate material transport, mixing, or size
segregation. Consequently, the fundamental properties and motions of
granular matter confined to a vibrating container have been widely
studied. Many fundamental studies have focused on the motion of
particle beds in a rectangular container (the pan).  Typically, the
bed is excited by pan vibrations along a single axis.  Reviews of such
studies can be found in
Refs.~\cite{JN92,BD92,M94,JNB96,JNBpt96,D98}. Agitated beds of
granular matter have been observed to adopt a wide spectrum of shapes
and motions, such as heaping, convection, small amplitude wave motion,
arching, and large amplitude wave motion; see
e.g. Ref.~\cite{WBH96}. These modes certainly depend on the nature of
the vibrations.  However, their dependence on the pan geometry is less
well understood.  For example, in the case of purely vertical
vibrations, the effect of the pan shape is absent or weak~\cite{Sxx},
but this is unlikely to be generally true in the presence of
horizontal vibrations.

\begin{figure}
\hbox to\hsize{\epsfxsize=1.0\hsize\hfil\epsfbox{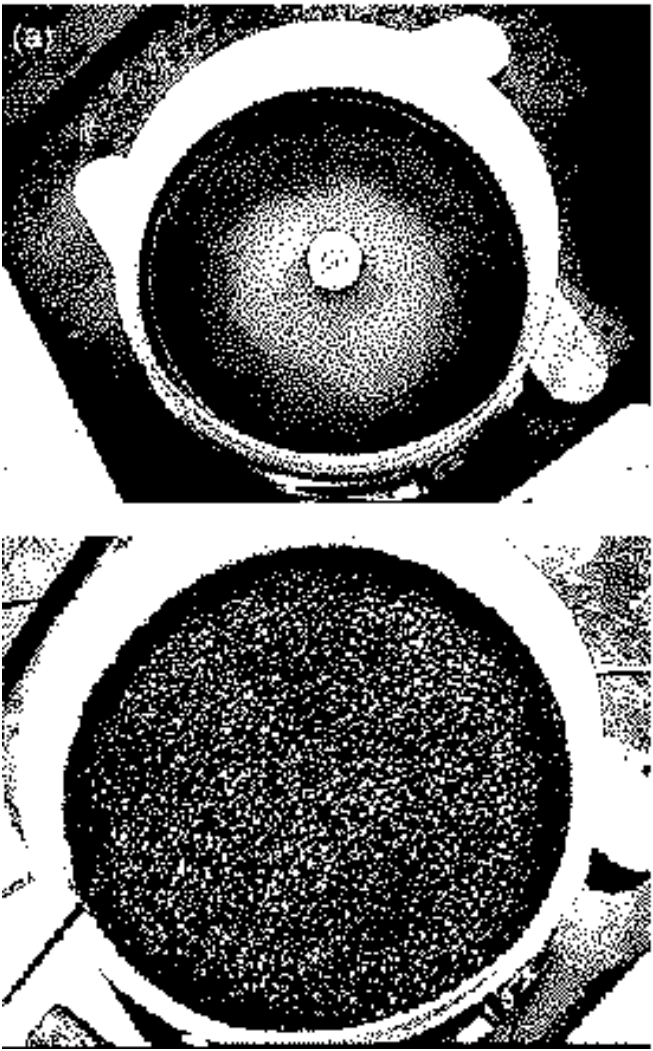}\hfil}
\caption{Photographs of the agitated bed apparatus, viewed from above:
(a) without particles; (b) with particles.}
\label{app1}
\end{figure}

The motions observed in a cylindrical pan subjected to simultaneous
horizontal and vertical vibrations, are reported in the present study.
There are two principal motivations for studying this case.  First,
the response of a bed of granular matter to excitations occurring
simultaneously along three Cartesian axes has not been extensively
studied and remains poorly understood.  However, in many industrial
applications, tri-axial excitations are highly desirable to optimize
process parameters.  Many fundamental studies have been made of
systems subjected to purely vertical or horizontal vibrations, but in
only a few cases are vibrations in both directions addressed. An
example is the work of Tennekoon and Behringer~\cite{TB98}, who
reported that when granular matter is subjected to simultaneous
horizontal and vertical sinusoidal vibrations, the phase difference
between the components of vibration in the two directions becomes a
key control parameter for the resulting motion.  A quantitative
understanding of the effect of such combined vibrations is therefore
important for predicting and controlling the behavior of agitated beds
in real applications.

A second motivation is to study the motion generated in a container of
cylindrical symmetry, with the cylinder axis oriented vertically.
Again, this geometry is common in industrial devices such as sifting
machinery, yet is less well studied in a research context compared to
square and rectangular geometries. Note that the present case is
distinct, both in motion and in orientation, from the more widely
studied case of a continuously rotating drum having the cylinder axis
oriented horizontally.

A cylindrical geometry provides the opportunity to study phenomena
unlikely to occur in the rectangular case.  In particular, the absence
of discontinuities (i.e. sharp corners) on the vertical boundaries of
a cylindrical pan makes possible smooth circulatory motions of the
agitated bed.  These circulation modes are of interest for
applications involving sifting and mixing, and for developing agitated
bed devices for continuous, as opposed to batch, processing.  Initial
studies have also demonstrated that these modes may display complex
behavior.  For example, Scherer et al.~\cite{SMER98} have reported
that when spheres are placed in a cylindrical pan and are subjected to
horizontal shaking, they circulate in one direction at low packing
densities, and in the other at high packing densities or high
excitation frequencies.

\begin{figure}
\hbox to\hsize{\epsfxsize=1.0\hsize\hfil\epsfbox{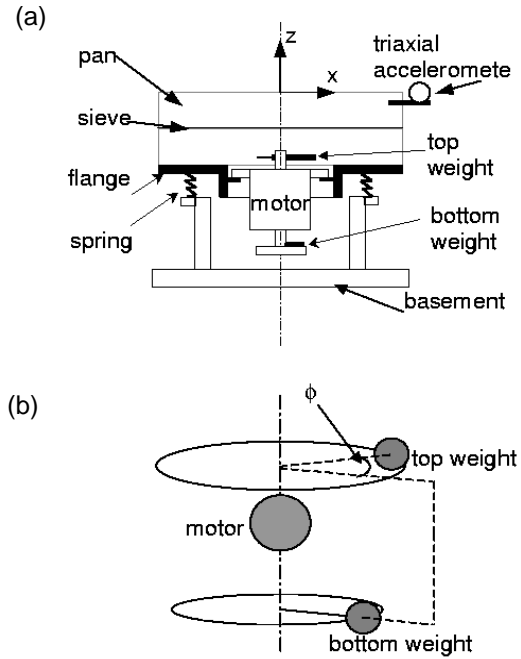}\hfil}
\caption{ (a) A schematic view of the apparatus.  (b) Schematic
representation of the relative positions of the motor and top and
bottom weights, and the definition of $\phi$.}
\label{app2}
\end{figure}

\section{EXPERIMENTAL SET-UP}
\label{expt}

The agitated bed apparatus studied here (Figs.~\ref{app1} and
\ref{app2}) is a modified SWECO finishing mill Model ZS30S66, a
commercially available device widely used in industrial applications
for sieving and polishing. It consists of a rigid, circular, stainless
steel pan of radius $R=0.381$~m with a flat bottom.  The pan
``floats'' on an array of nine springs spaced equidistantly around the
circumference of the pan bottom. Vibrations are excited by a $0.5$~HP
motor mounted below the center of the pan.

In order to induce simultaneous horizontal and vertical vibrations,
and concurrently obtain control over their relative magnitudes, metal
weights are mounted eccentrically at the top and bottom ends of a
shaft running through the center of the motor. The motor housing is
rigidly attached to the pan via a flange, while the shaft rotates
freely with respect to the pan.  When the shaft rotates, the effect of
the weights is to impose an oscillatory torque that tends to make the
motor shaft deviate from the vertical.  The affect of this torque is
to cause the center of the pan to revolve about a vertical axis, and
simultaneously cause the bottom of the pan to deviate from the
horizontal. A complete description of the pan motion is given in the
next section. The angle between the top and bottom weights, referred
to here as the ``lead angle'' $\phi$, can be varied from $0^\circ$ to
$180^\circ$ [Fig.~\ref{app2}(b)]. The motor shaft rotates the weights
in a range of frequency $\omega$ of $10$ to $20$~Hz.

The bottom surface of the pan is a stiff, stainless steel seive. The
particles used in this study are too large to fall through the holes
of the seive.  The springs that bear the load of the entire pan-motor
assembly are mounted on a rigid basement.  Note also that there is a
cylindrical ``hub'' attached to the center of the pan bottom, of
radius 10.4~cm and height 3.0~cm.

The masses of the top and bottom weights attached to the motor are
fixed in this study.  As will be shown in the next section, the angle
$\phi$ between the weights controls the ratio of the amplitudes of the
horizontal and vertical vibrations imparted to the pan. The rotational
frequency $\omega$ of the motor controls the amount of energy
introduced into the system.  The principal control parameters for the
experiments described here are thus $\phi$ and $\omega$.

The experiments are conducted with two sets particles: ``pill-shaped''
oblate spheroid particles, and spherical particles. The oblate
spheroids have a major axis of 13.5~mm and a minor axis of 7~mm, a
mass of $0.9$~g and a density of 1200~kg/m$^3$, with a hard and smooth
polished surface. The spheres have the same volume as the oblate
spheroids, a density of 1100~kg/m$^3$, and a hard, smooth but
unpolished surface. Although the qualitative behavior observed in both
cases is similar, it is found that the shape of the particles
influences some quantitative properties of the motion.  Thus the
results obtained with each type of particle are presented separately.

The mass of the unloaded pan is approximately 150~kg.  All tests are
conducted for particle bed loads below 45~kg and it is found that the
influence of the bed load on the pan motion is negligible.  Also, it
is confirmed in all tests that the pan moves as a rigid body.

The motion of the pan is monitored using Entran accelerometers: one
tri-axial (model EGA3-F-10-/5) and three mono-axial (model
EGA-F-10-/5) units are used. The location of the tri-axial
accelerometer is shown in Fig.~\ref{app2}; the others are positioned
at various locations along the top edge of the pan, according to the
degree of freedom to be measured.  The accelerometer data are
collected using a computer-based data acquisition system and processed
to obtain the velocity, displacement and phase-shift data for the pan
motion. A sampling frequency of 2~kHz is used for data acquisition,
allowing a minimum resolution of 165 points/cycle. This results in an
error in the phase angle measurements of $\pm 3^\circ$.  The
uncertainty in the acceleration and bed displacement are estimated to
be $\pm 0.1$~g and $\pm 0.05$~mm, respectively, where $g$ is the
acceleration due to gravity.

The surface motion of the particle bed is studied by visually tracking
particles.  Differently colored particles are used to obtain
information on the average mixing time. For the most interesting modes
of motion, a localized group of particles is coated with a fluorescent
dye and the subsequent motion of this group is recorded
photographically.  The dye is sodium fluorescene, which is excited
using an ultra-violet light source and observed in the 532~nm visual
range.

The results of the experiments are presented in two stages.  First, a
quantification of the pan motion is presented in Section 3.  Then the
modes of particle motion that are observed in the bed are described in
Section 4.

\section{PAN MOTION}

As described above, the experimental apparatus is a commercially
available unit that generates simultaneous horizontal and vertical
vibrations of the pan.  The first task of the present study is
therefore to quantify the motion generated by this device.  To achieve
this, two coordinate systems are defined: one fixed with respect to
the pan (the ``pan frame''), and one fixed with respect to the
laboratory (the ``lab frame'').

The intersection of the cylinder axis of the pan with the plane
defined the by top edge of the pan is chosen as the origin of the
Cartesian coordinate system $(x_p,y_p,z_p)$ fixed in the pan frame.
The cylindrical symmetry of the pan is broken by the existence of a
spout located at a point on the circumference of the pan.  The
$x_p$-axis in the pan frame is thus defined as the line that passes
through the origin and the location on the spout where the triaxial
accelerometer is mounted.  The $y_p$-axis is the line perpendicular to
the $x_p$-axis in the plane defined by the top edge of the pan, and
the $z_p$-axis is perpendicular to both the $x_p$ and $y_p$-axes.  The
Cartesian coordinate system $(x,y,z)$ in the lab frame is defined as
that which is coincident with the pan frame when the apparatus is at
rest.

\begin{figure}
\hbox to\hsize{\epsfxsize=1.0\hsize \hfil\epsfbox{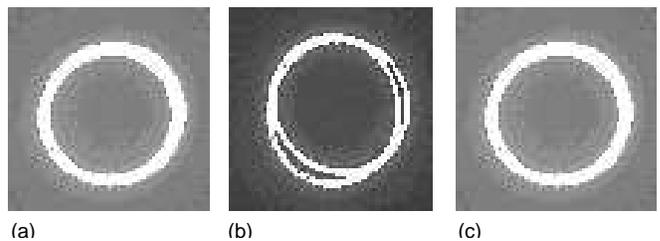}\hfil}
\caption{Photographs of laser traces visualizing the circular motion
of the pan center for (a) $\phi=10^\circ$, $\omega=30$~Hz; (b)
$\phi=60^\circ$, $\omega=30$~Hz; and (c) $\phi=100^\circ$,
$\omega=30$~Hz.}
\label{laser}
\end{figure}

The accelerometers provide data that are analyzed to give the lab
frame coordinates as a function of time for points fixed in the pan
frame.  To specify the motion of the pan, in the following attention
is restricted to points fixed in the pan frame lying in the $z_p=0$
plane, since this is the plane in which the accelerometers are
located.  Based on the accelerometer data, the motion in the lab frame
of such a point is found be consistent, within the error of
measurement, with the following model:
\begin{eqnarray}
x(t)&=&x_p+x_{\rm max}\cos(\omega t) \label{x} \\ 
y(t)&=&y_p+y_{\rm max}\cos(\omega t+\alpha_{xy}) \label{y} \\ 
z(t)&=&A(r_p)\cos(\omega t+\alpha_{xz}-\theta), \label{z}
\end{eqnarray} 
where,
\begin{eqnarray}
A(r_p)&=&\frac{r_p}{R}z_{\rm max},
\label{A}
\end{eqnarray}  
and
\begin{eqnarray}
r_p&=&\sqrt{x_p^2+y_p^2}.
\label{r} 
\end{eqnarray}  
In the above equations, $x_{\rm max}$ and $y_{\rm max}$ are the
amplitudes of the $x$ and $y$ displacements of the pan-frame origin in
the lab frame; $\alpha_{xy}$ is the phase shift between the $x$ and
$y$ displacements in the lab frame; and $\theta$ is the angle measured
about the origin from the $x_p$-axis to the point $(x_p,y_p)$.
$\alpha_{xz}$ is the phase shift between the $x$ and $z$ displacements
in the lab frame, as measured on the $x_p$-axis; thus
$(\alpha_{xz}-\theta)$ is the phase shift between the $x$ and $z$
displacements in the lab frame, to be found at the point $(x_p,y_p)$.
$A(r_p)$ is the amplitude of the $z$ displacement of a point in the
$z_p=0$ plane that is a distance $r_p$ from the origin in the pan
frame.  $R=0.381$~m is the distance from the origin in the pan frame
to the edge of the pan, and $z_{\rm max}$ is the amplitude of the $z$
displacement of a point on the edge of the pan (i.e. at $r_p=R$).

\begin{figure}
\hbox to\hsize{\epsfxsize=1.0\hsize\hfil\epsfbox{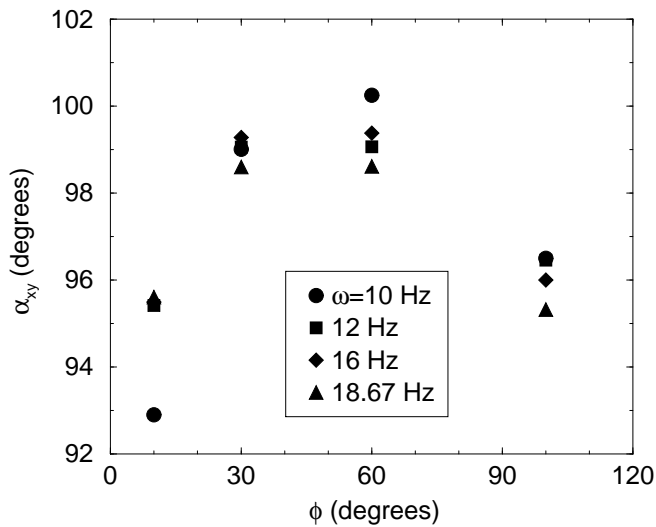}\hfil}
\caption{Phase shift $\alpha_{xy}$ as a function of $\phi$ for various
$\omega$}
\label{axy}
\end{figure}

The form of Eqs.~\ref{x} and \ref{y} is motivated by the observation
that under all operating conditions, the center of the pan appears to
execute an approximately circular orbit in a horizontal plane.  To
visualize this, a laser source is mounted at the center of the pan,
pointing upward, normal to the pan bottom.  The motion of the pan
causes the laser to trace out a pattern on a screen positioned
horizontally above the apparatus.  Time-exposure photographs of these
traces are shown in Fig.~\ref{laser} for various operating conditions,
and are consistent with circular motion of the pan center.

\begin{figure}
\hbox to\hsize{\epsfxsize=1.0\hsize\hfil\epsfbox{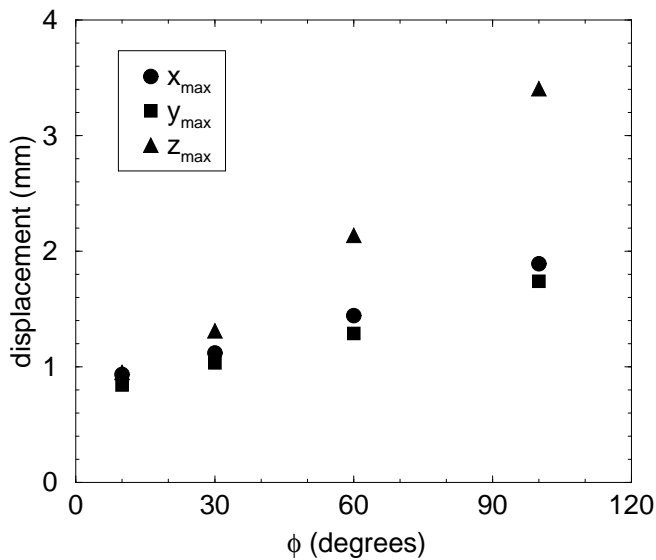}\hfil}
\caption{Maximum displacements $x_{\rm max}$, $y_{\rm max}$ and
$z_{\rm max}$ as a function of $\phi$.  These data are averages over
approximately 20 runs over the range of $\omega$ studied.  No
systematic variation is observed as a function of $\omega$, though
individual measurements are scattered about the average by about
$15\%$.}
\label{max}
\end{figure}

\begin{figure}
\hbox to\hsize{\epsfxsize=1.0\hsize\hfil\epsfbox{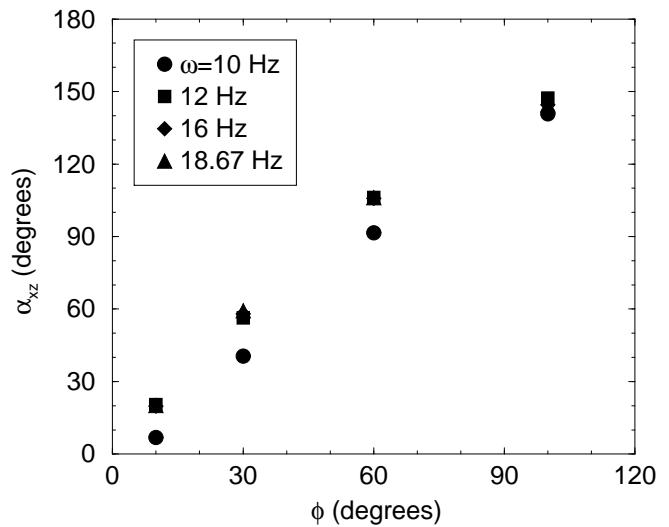}\hfil}
\caption{Phase shift $\alpha_{xz}$, as a function of $\phi$ for
various $\omega$.}
\label{axz}
\end{figure}

If the motion is nearly circular, the magnitude of the phase shift
$\alpha_{xy}$ should be approximately $90^\circ$.  This is confirmed
in Fig.~\ref{axy}, where $\alpha_{xy}$ is plotted as a function of
$\phi$ for various $\omega$.  An apparatus with perfect cylindrical
symmetry would have $\alpha_{xy}=90^\circ$.  Small asymmetries in the
apparatus, the most important of which is likely the spout on the edge
of the pan, result in the observed variation in the range
$93^\circ<\alpha_{xy}<100^\circ$.

Circular motion of the pan center also requires $x_{\rm max}$ and
$y_{\rm max}$ to be equal.  Measurements of $x_{\rm max}$ and $y_{\rm
max}$ at $r_p=R$ (Fig.~\ref{max}) show that this is observed to be
true within the measurement error across the range of $\phi$ studied
here.  Note that as $\omega$ varies in different test runs, $x_{\rm
max}$ and $y_{\rm max}$ are found to be scattered around the average
values reported in Fig.~\ref{max} by about $15\%$; however, no
systematic variation with $\omega$ can be identified.  Tests also show
that the values of $x_{\rm max}$ and $y_{\rm max}$ are independent of
the position around the edge of the pan at $r_p=R$ at which they are
measured, consistent with the description in Eqs.~\ref{x} and \ref{y}.
Note that for $x_{\rm max}$ or $y_{\rm max}$ to have the same measured
value independent of $r_p$ at the point of measurement requires that
the tilt angle of the $z_p=0$ plane with respect to the $z=0$ plane be
small.  This is confirmed below.

Eq.~\ref{z} characterizes the tilting motion of the pan by quantifying
the vertical deviation of a point fixed in the pan frame (in the
$z_p=0$ plane) with respect to the $z=0$ plane in the lab frame.  As
in the case of the $x$ and $y$ motions, the accelerometer data for
$z(t)$ indicate a sinusoidal function of $t$.  The form of Eq.~\ref{z}
also assumes that the origin of the pan frame does not leave the $z=0$
plane of the lab frame; this assumption is confirmed by direct visual
observation using a high-speed (500 frames/s) camera.  From geometric
considerations, the amplitude $A$ of the $z$ variation in Eq.~\ref{z}
is a function of the distance $r_p$ of the point of measurement from
the center of the pan, as indicated in Eq.~\ref{A}.  The phase shift
between the $x$ and $z$ displacements is a difference of two
contributions, $\alpha_{xz}$ and $\theta$.  $\alpha_{xz}$ is the phase
shift between the $x$ and $z$ displacements as measured on the $x_p$
axis at $r_p=R$.  $\theta$ is a geometrical term that accounts for the
difference between the phase shift observed on the $x_p$ axis and that
which would be observed at an arbitrary point $(x_p,y_p)$ in the
$z_p=0$ plane.  As shown in Fig.~\ref{axz}, $\alpha_{xz}$ does not
depend strongly on $\omega$, but does depend on $\phi$.

\begin{figure}
\hbox to\hsize{\epsfxsize=1.0\hsize\hfil\epsfbox{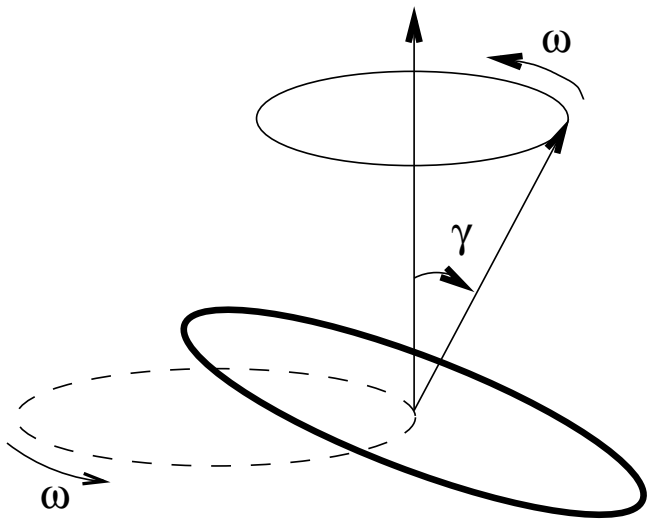}\hfil}
\caption{Schematic representation of the pan motion.  The dashed
circle represents the orbit of the pan centre in the lab frame.  The
solid circle represents the rim of the pan.}
\label{gamma}
\end{figure}

The net effect of the motions described by Eqs.~(1-3) are summarized
in Fig.~\ref{gamma}.  The motion of the pan can be thought of as a
superposition of two motions:

(i) The center of the pan revolves about the lab frame origin in a
nearly circular, horizontal orbit.  This motion is directly described
by Eqs. (1-2).

(ii) The pan is tilted at an angle $\gamma$ in such a way that a unit
normal vector rooted at the pan center precesses about the vertical at
the same frequency at which the center orbits the origin, but phase
shifted with respect to the orbital motion of the center.  The fact
that $\gamma$ maintains a constant value can be shown from Eq. (3) by
considering (e.g.) the case $r_p=R$ and setting $\theta=\omega
t+\alpha_{xz}$.  These conditions describe the point on the edge of
the pan that at any given $t$ has the largest positive $z$
displacement.  For all $t$, this displacement is a constant, $z_{\rm
max}$, and so $\gamma=\sin^{-1}(z_{\rm max}/R)$ is also constant.  Note
that for the apparatus and operating conditions described here,
$R=0.381$~m, and $z_{\rm max}$ is never greater than 0.0035~m, giving
$\gamma<0.5^\circ$.  This confirms the assumption of small tilt angle
required for the chosen form of Eqs. (1-2), as indicated above.

The measurements shown in Figs.~\ref{axy}-\ref{axz} show that both the
amplitudes and phase shifts required to specify the pan motion using
Eqs. (1-3) are mainly controlled by $\phi$.  The effect of $\omega$ is
less problematic, since the ratio of displacements $x_{\rm max}$ and
$z_{\rm max}$ do not vary in a systematic way with $\omega$.  However,
since $\omega$ controls the rate at which energy is introduced into
the system, both $\phi$ and $\omega$ will play a major role in
determining the particle motion, as shown next.

\begin{figure}
\hbox to\hsize{\epsfxsize=1.0\hsize\hfil\epsfbox{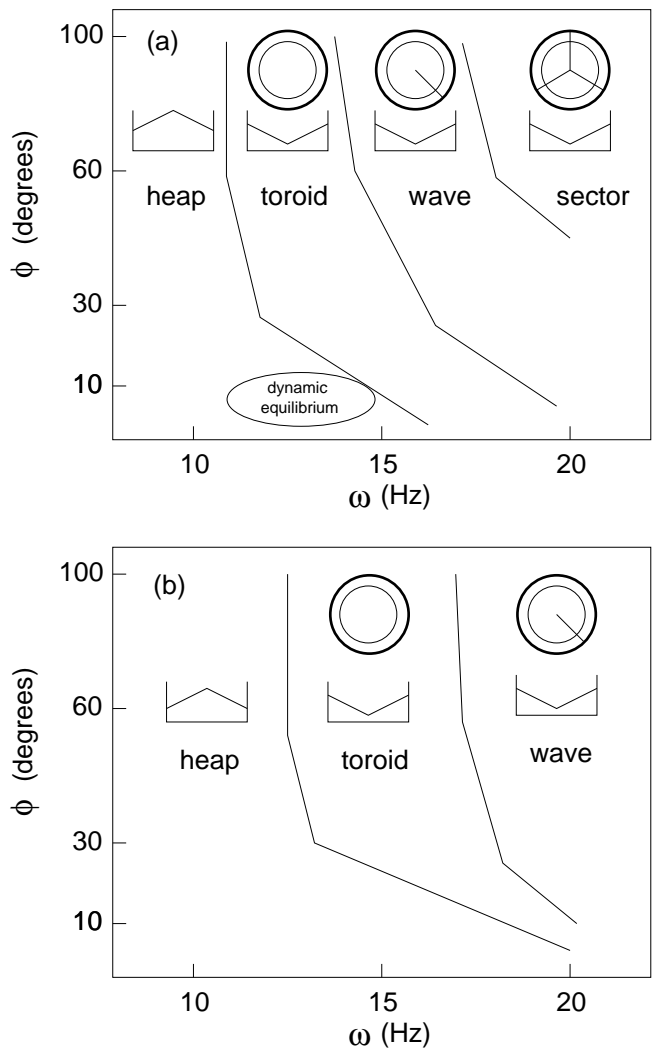}\hfil}
\caption{Diagram identifying the regions of the $\omega$-$\phi$ plane
in which distinct bulk dynamical modes of the particle bed are
observed for (a) oblate spheroids, and (b) spheres. Note that in (a) a
region of dynamic equilibrium is observed near $\phi = 10^\circ$,
where the bed does not perform any net rotation about the pan centre.
This is the boundary region between counter-clockwise and clockwise
bed rotation (``laps'').}
\label{pd}
\end{figure}

\section{MODES OF PARTICLE BED MOTION}

A diagram showing the observed bulk dynamical modes of the particle
bed as a function of $\omega$ and $\phi$ is shown in Fig.~\ref{pd}(a)
for the oblate spheroids, and in Fig.~\ref{pd}(b) for spheres.  The
bulk motion of the bed is a strong function of the $\omega$.  Four
distinct modes of bed motion are observed, here termed ``heaping,''
``toroidal motion,'' ``surface waves'' and ``sectors.''  Each of these
is described in detail below. The range of the $\omega$ at which each
of the modes appears depends on the particle shape as can be seen from
Fig.~\ref{pd}, but the qualitative behavior of the modes is mostly
similar. In the following sections, the motion in general is
discussed, and differences due to particle shape are identified where
appropriate.

\subsection{Heaping}

\begin{figure}
\hbox to\hsize{\epsfxsize=1.0\hsize\hfil\epsfbox{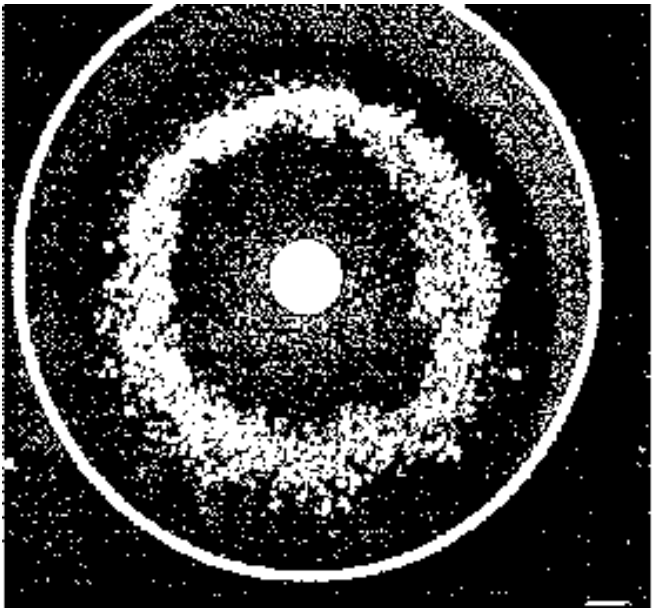}\hfil}
\caption{Fluorescene dye visualization of the particle bed surface
motion in the heaping mode.  30~kg, $\phi=60^\circ$, $\omega=30$~Hz.}
\label{heap}
\end{figure}

At the lowest $\omega$ studied the particles form a heap with an
elevated center. The relative particle movement and mixing is very
small and the particle bed rotates almost as a solid body.  This
motion is visualized in Fig.~\ref{heap}.  Fluorescent dye is
continuously added at a specific location to the surface of the
particle bed while in steady-state motion.  The subsequent motion of
the dyed particles is then captured photographically under
ultra-violet light, as shown in Fig.~\ref{heap}.  The dyed particles
on average undergo a slow processional motion about the pan center.
The radial component of the particle motion and the diffusion of
particles is negligible.  Notably, for most $\omega$ and $\phi$, the
particle bed rotates in the direction opposite to that of the orbital
motion of the center of the pan; the exceptions to this occur at the
lowest $\phi$ and $\omega$ studied (e.g. $\phi=10^\circ$ and 16~Hz for
oblate spheroids), where the rotational directions are the same.  The
direction of motion of the bed appears to depend on the phase
difference $\alpha_{yz}= \alpha_{xz} - \alpha_{xy}$.  For $\alpha_{yz}
> 0^\circ$, the bed rotates in the direction opposite of the pan
center.
   
The slope of the particle bed (from the center to the outer edge)
increases with an increase in $\phi$. For example, the bed of oblate
spheroids attains the maximum slope for $\phi=100^\circ$, and a nearly
flat surface when $\phi=10^\circ$.  Fig.~\ref{pd} schematically shows
the cross-sectional shape of the surface during the heaping mode.

The spherical particles move in more tightly packed layers than the
oblate spheroids. For $\phi=100^\circ$, the layers are very tightly
packed and the local arrangement of particles is similar to a
close-packed lattice.  As $\phi$ increases, the uniformity of the
lattice breaks down.  The bed is still tightly packed at $\phi=
100^\circ$, but more pronounced relative motion of the particles is
observed.  This change in behavior may be related to the increase of
$z_{\rm max}$ as $\phi$ increases (see Fig.~\ref{max}).

Previous studies of granular matter in vertically excited rectangular
containers~\cite{LR97} have shown that relative particle motion is
observed when the maximum vertical acceleration exceeds $1.2g$.  In
the present tests, it is found that the local acceleration is below
this threshold for $\omega<10$~Hz.  However, at $\phi = 100^\circ$,
the acceleration at the edge of the pan does approach $1.2g$.

\subsection{Toroidal Motion}

\begin{figure}
\hbox to\hsize{\epsfxsize=1.0\hsize\hfil\epsfbox{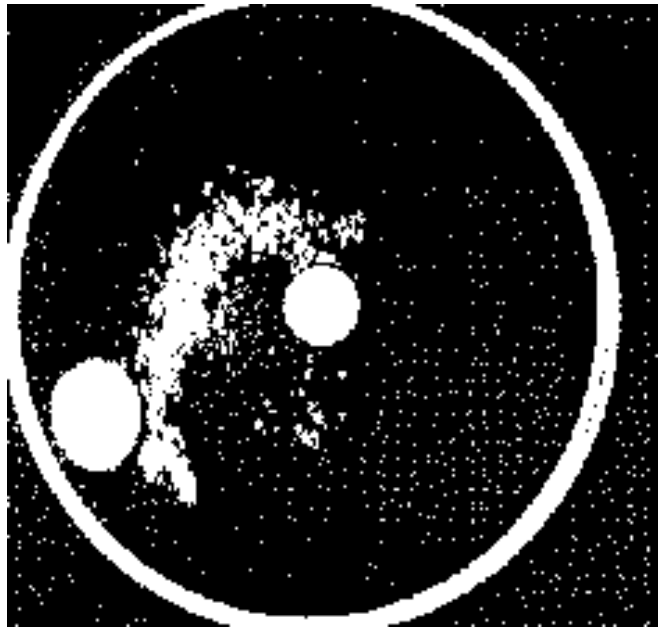}\hfil}
\caption{Fluorescene dye visualization of the particle bed surface
motion in the toroidal mode.  Particles spiral to the center of the
bed along the surface and toward the edges along the pan surface.
These reappear at the surface of the bed along the edges. 30~kg,
$\phi=60^\circ$, $\omega=56$~Hz.}
\label{dt}
\end{figure}

\begin{figure}
\hbox to\hsize{\epsfxsize=1.0\hsize\hfil\epsfbox{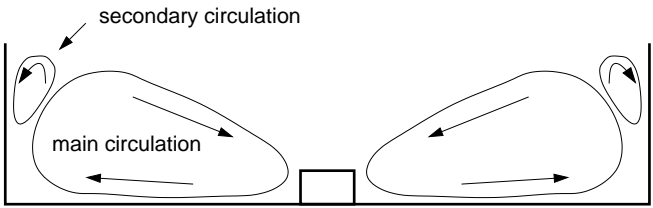}\hfil}
\caption{Schematic of cross-section of particle bed, showing direction
of particle flows in toroidal motion.}
\label{dtschem}
\end{figure}

As $\omega$ increases the particle bed undergoes a transformation to a
toroidal shape and the particle motion becomes organized.  This motion
is highly coherent as seen in the fluorescent dye visualizations in
Fig.~\ref{dt}.  Dye is continuously injected onto the surface of the
moving bed.  Particles initially on the outer edge move in a spiral
motion to the center.  Near the center they are subducted and travel
outward along the bottom of the pan, to be re-entrained along the
walls and reappear at the surface at the outer edge of the pan.  As
seen for the heap, the bulk rotational motion of the bed in the
horizontal plane is opposite to that of the center of the pan.  The
entire particle bed adopts the shape of a torus, through which
individual particles move in helical trajectories.  The particle
motion in a cross-sectional plane through the toroid is depicted
schematically in Fig.~\ref{dtschem}.  Note that in addition to the
primary circulation roll occupying the center portion of the bed, a
small secondary (and counter-rotating) circulation roll forms on the
top of the primary one, near the pan wall. The particles move toward
the center along the top surface in the primary roll, and toward the
wall along the top surface in the secondary roll. The slope of the
particle bed surface is greater in magnitude, and opposite in sign,
than that observed for the heap motion.

In this mode of motion, the particle bed is fluidized.  The increased
agitation results in a relatively rapid diffusion of the particles and
a net increase in the mixing rate.

The toroidal motion observed for the circular pan has not been
observed in rectangular geometries.  An inward spiraling motion has
been observed for granular media underneath rotating
fluids~\cite{SMER98} and the radial segregation of granular mixtures
has been observed in rotating cylinders~\cite{KMOXX}.  For vertically
excited rectangular beds, stationary convection cells, similar to
Rayleigh-Benard instabilities, are observed~\cite{KEKFJN96,LR97}.  For
horizontal vibrations, experimental~\cite{TKB98} and
computational~\cite{SP01} studies have shown that the convective
motion consists of granular matter rising to the surface at the middle
of the container.  In contrast, in the present study, the particles
rise along the walls and are subducted at the middle of the pan.

\subsection{Surface Waves}

For the two modes described above, the bed height is only a function
of the distance from the center of the pan, and the shape of the bed
is symmetric about the cylinder axis of the pan.  However, for
$\omega>15$~Hz, crests and troughs appear on the surface of the bed as
a function angular direction around the pan.  The crests, or fronts,
propagate in the same direction but at a higher speed than the bulk
rotation of the bed in the horizontal plane. The speed of the particle
fronts initially increases as $\omega$ increases, but then decreases
as $\omega$ approaches $18$~Hz.  The amplitude (i.e. the height of the
crest above the bed) decreases as $\omega$ increases. Unfortunately,
this motion is very difficult to capture on still media.
 
The individual particle trajectories can still be described as
helical.  The motion of the crests appears to be similar to a
traveling wave.
 
\begin{figure}
\hbox to\hsize{\epsfxsize=1.0\hsize\hfil\epsfbox{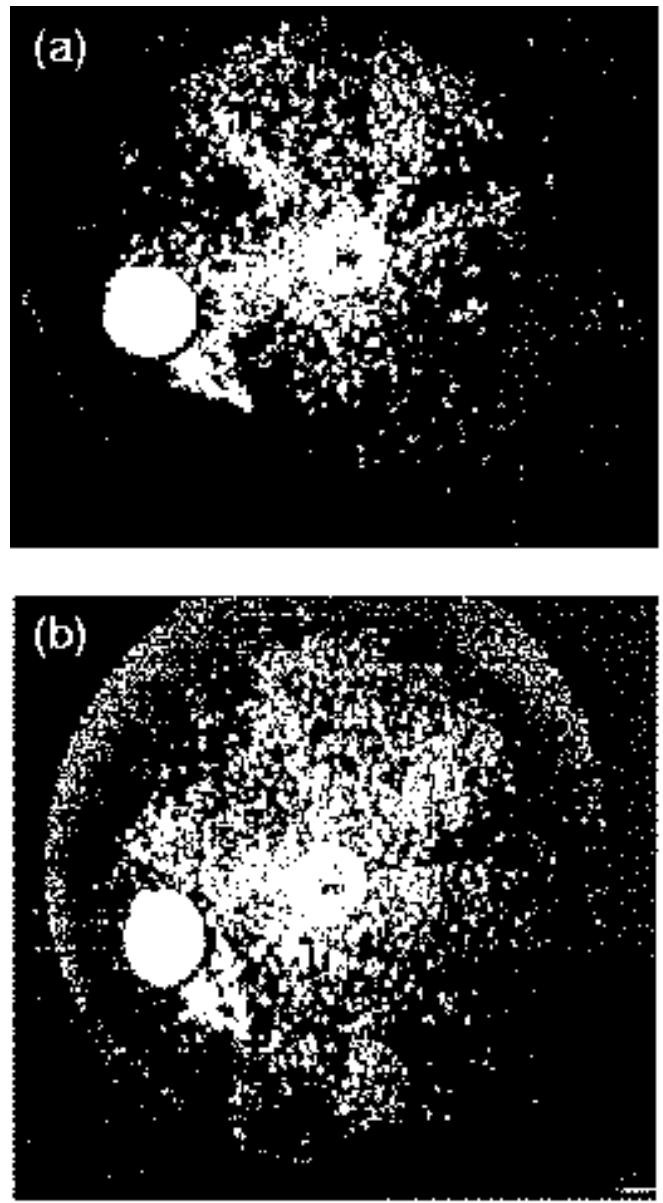}\hfil}
\caption{Two fluorescene dye visualizations of sectors.  The sectors
can be visualized because the particles located on the crests dry
faster than those in the trough and loose fluorescence.}
\label{sector}
\end{figure}

\subsection{Sectors}

For $18$~Hz$<\omega<20$~Hz, the motion of crests on the surface of the
bed becomes nearly stationary, dividing the bed into well-defined
``sectors.'' These sectors are visualized using fluorescent dye in
Fig.~\ref{sector}. The number of sectors present in the bed is found
to be a function of $\phi$ and $\omega$.  The sectors are stable and
easily recognized for the oblate spheroids.  However, for the
spherical particles, stationary sectors are more difficult to achieve
and could not be reliably reproduced.
     
Although the general bulk motion of the particles remains helical, the
sector mode is distinct from the toroidal mode.  Defining the general
direction of the processional motion as streamwise, it is seen that
the particles on the upstream side of the crests are drawn into the
bed, while on the downstream side particles emerge on the surface.
The upstream particles have been exposed to air for a longer period
and have dried, causing the dye to loose florescence.  The upstream
side of the crests thus appear darker than the downstream side,
yielding the ``rays'' seen in Fig.~\ref{sector}.  Also, in the sector
mode, the secondary circulation roll near the wall that was
characteristic of the unmodified toroidal mode of motion could not be
observed.

As $\omega$ is further increased to 26~Hz, the toroidal motion
(without sectors or surface waves) is recovered.  Unfortunately, the
mechanical limitations of the apparatus did not allow testing at
larger $\omega$.

\section{DISCUSSION}

\subsection{Rotation of the particle bed in the horizontal plane}

The direction of rotation of the particle bed in the horizontal plane
is generally opposite to that of the center of the pan.  This behavior
is initially counter-intuitive, and explaining it requires a detailed
understanding of the interplay between the bed motion, the pan
orientation, and the horizontal and vertical components of the pan
motion.

A crucial element for resolving this behavior is to determine the
shape and motion of the zone of contact of the particle bed with the
pan.  Because the pan is tilted away from the vertical at all times
during its motion, it seems likely that only a particular sector of
the bed is in contact with the pan at any given time; and, that the
angle at which maximum contact occurs between the bed and the pan
rotates about the pan center at the same frequency $\omega$ as the pan
center itself.  If this is true, at least two contributions to the
mechanism of bed rotation can be envisioned:

\begin{figure}
\hbox to\hsize{\epsfxsize=1.0\hsize\hfil\epsfbox{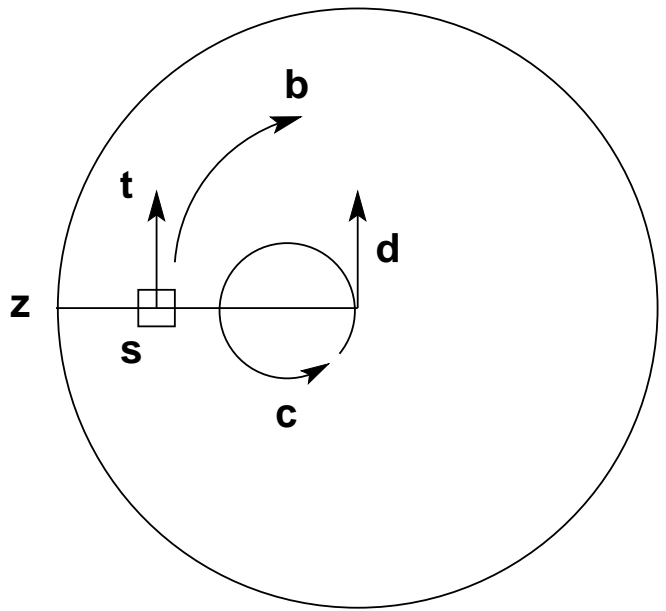}\hfil}
\caption{ Schematic illustration of scenarios for explaining direction
of bed rotation.  The large circle represents the edge of the pan as
viewed from above. $c$ is the circular trajectory of the centre of the
pan as seen in the lab frame, while $d$ is the instantaneous direction
of motion of the pan centre at the moment depicted.  $s$ is a surface
element of the pan bottom along the direction $z$ of the zone of
maximum contact between the pan and the bed.  Scenario (i): If the
vector $t$ represents the tangential component of the normal to $s$,
then particles will be deflected along $t$, leading to bed rotation in
the direction $b$, opposite to $c$.  Scenario (ii): If the vector $t$
represents the tangential component of the velocity of $s$, then
particles will be ``dragged'' in the direction of $t$, leading to bed
rotation along $b$, again opposite to $c$.}
\label{spec}
\end{figure}

(i) {\it Deflection by the pan surface:} Consider an element of the
pan surface within the zone of contact between the pan and bed.  At
any given time, the normal to this suface element can be decomposed
into vertical, radial, and tangential components, defined with respect
to a cylindrical coordinate system in the lab frame.  Particles
striking this surface element will tend to be deflected according the
to the orientation of the surface normal.  In the absence of other
mechanisms, a net rotational motion about the pan centre will occur if
the direction of the tangential component remains constant.  This will
be true if (as assumed above) the zone of particle-bed contact rotates
around the pan bottom at the same frequency as the pan itself
precesses.  The direction of this bed rotation will be independent of
the rotational direction of the pan center, since it depends on the
(presently unknown) location of the region of particle-bed
contact. Hence, the bed could well be set into a rotational motion
opposite to that of the pan center (Fig.~\ref{spec}).

(ii) {\it Entrainment by the pan surface}: Independent of the above
mechanism, the effect of friction between the particles and the pan
bottom should also be considered.  As described in Section~\ref{expt},
the pan bottom consists of a metal screen, and so is ``rough'' on the
scale of the particles themselves. Consider the same surface element
in the contact zone as discussed above, but instead of its
orientation, consider its velocity vector decomposed into vertical,
radial, and tangential components.  This rough surface element will
tend to transfer momemtum to particles in contact with it according to
the direction of its velocity.  Analogous to the reasoning given
above, particles in the contact zone will be subjected to a constant
tangential surface velocity, yielding a net rotation of the bed about
the pan center.  Also as above, the direction of the resulting bed
rotation will be independent of that of the pan center, and so it
should be possible to establish conditions where these rotations would
be in opposite in direction (Fig.~\ref{spec}).

Whatever the underlying mechanism, an explanation can be tested
against the observed dependence of the motion on the control
parameters.  For example, the speed of rotation of the bed in the
horizontal plane depends on the phase difference between the
horizontal and vertical pan motions, $\alpha_{xz}$.  For $0^\circ<
\alpha_{xz} < 180^\circ$, the particles are observed to move in the
direction opposite to that of the bed.  However, as mentioned earlier,
it is indeed possible to realize conditions where the bed rotates in
the same direction as the pan, when $\phi$ is changed such that
$\alpha_{xz}<0^\circ$.

\begin{figure}
\hbox to\hsize{\epsfxsize=1.0\hsize\hfil\epsfbox{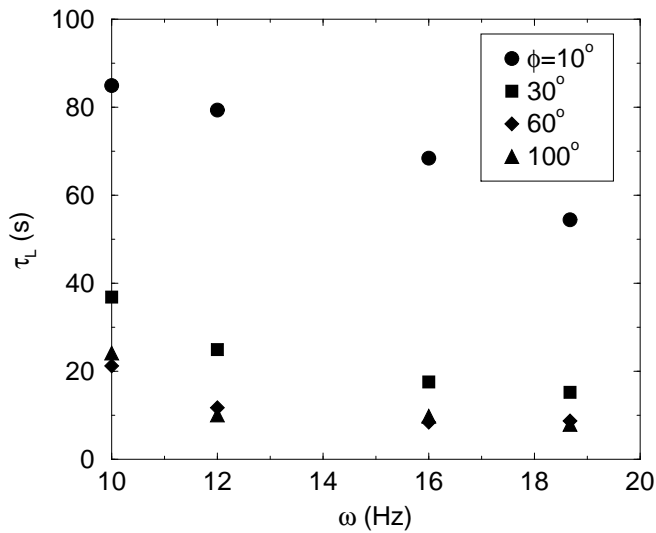}\hfil}
\caption{A plot of the average lap time $\tau_L$ in seconds.}
\label{tl}
\end{figure}

In addition, the rotational speed of the bed in the horizontal plane
increases with $\omega$.  Furthermore, as shown in Fig.~\ref{tl}, the
rotational speed is found to be greatest when $\alpha_{xz} =
90^\circ$, corresponding to $\phi = 60^\circ$ (see Fig.~\ref{axz}).

Elucidation of these phenomena requires information on the
(time-dependent) location of contact between the pan and the bed.
This information is not available in the present measurements.  These
questions are being explored through further experiments, and through
computer modelling.

\subsection{Radial motion of the particles}

\begin{figure}
\hbox to\hsize{\epsfxsize=1.0\hsize\hfil\epsfbox{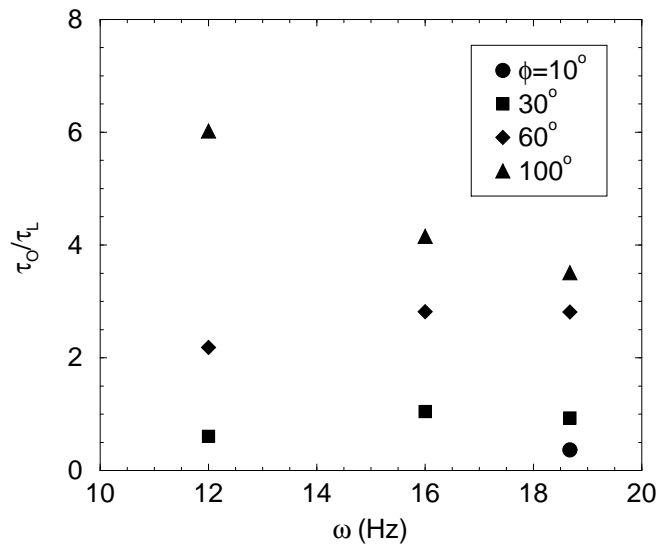}\hfil}
\caption{A plot of the overs to lap ratio for the oblate spheroids.  }
\label{tot}
\end{figure}

Understanding the radial motion of the particle bed also requires a
careful consideration of the bed and pan motions. In the toroidal
mode, the particles move along the bottom of the bed toward the wall,
and then back toward the center of the pan along the top surface of
the bed.  It seems likely that the motion of particles near the pan
bottom is due to their being entrained by the motion of the pan
surface itself.  The inward motion of particles on the bed surface,
where the particle packing is looser, may be dominated by
gravity-induced downward flow toward the bed center, where the bed
depth is smallest.  However, these possible explanations do not fully
elucidate the origin of the toroidal shape adopted by the bed in this
mode, and so further study of this behavior is needed.

The ratio of the rotational speed of the particle bed in the
horizontal plane, to the radial speed is show in Fig.~\ref{tot}.  To
measure this, the motion of color-tagged particles is observed in
steady state.  The average time $\tau_L$ required for a tagged
particle to complete one circuit (a ``lap'') around the pan in the
horizontal plane is measured; the average time $\tau_O$ required for a
tagged particle to complete one circuit (an ``over'') from the center of
the pan to the outside edge and back to the center is also measured.
The ratio $\tau_L/\tau_O$ is shown in Fig.~\ref{tot}.
  
In the heaping mode the radial motion is negligible, as indicated in
Fig.~\ref{heap}.  When the vertical acceleration is sufficiently high
to fluidize the particles (i.e. the particles are not in continuous
contact with the pan surface), the radial and horizontal rotational
speeds of the particles are found to depend only on the phase
difference between the horizontal and vertical pan motions.  This is
shown in Fig.~\ref{tot}: as $\omega$ is increased, $\tau_O/\tau_L$
approaches a constant value, which is a function of $\phi$.
    
\subsection{Critical acceleration and fluidization}

\begin{figure}
\hbox to\hsize{\epsfxsize=1.0\hsize\hfil\epsfbox{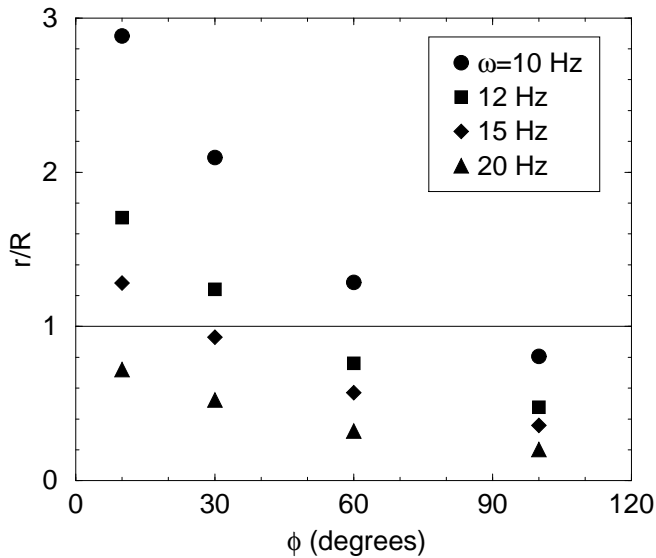}\hfil}
\caption{Critical radius at which the vertical acceleration is $g$ as
a function of $\phi$.}
\label{roR}
\end{figure}

The data in Fig.~\ref{max} are used to determine the critical radius,
shown in Fig.~\ref{roR}, at which the maximum vertical acceleration
exceeds $g$.  For $\phi = 10^\circ$ and $\omega<18$~Hz, the vertical
acceleration is less than $g$ and the particles at the bottom of the
bed are constantly in contact with the pan surface.  This is
consistent with the radial motion of the particle bed being
negligible, as indicated in Fig.~\ref{pd}.  Based on the model motion
in Eqs.~\ref{x}-\ref{z}, for $\omega=20$~Hz, the vertical acceleration
exceeds $g$ when $r_p/R > 0.81$.  Under these conditions the toroidal
motion is induced, in which case an overs-to-lap ratio could be
clearly defined (Fig.~\ref{tot}).  As $\phi$ increases, the critical
value of $g$ occurs in the bed at smaller $\omega$, in agreement with
the observations summarized in Fig.~\ref{pd}.

When the maximum vertical acceleration in the pan is less than $g$,
the tilting motion of the pan induces only a rotational motion of the
bed in the horizontal plane, since the particles at the bed bottom
remain more or less in constant contact with the pan surface and the
net radial bed motion is zero.  When the maximum acceleration exceeds
$g$, the particles at the bottom of the bed are lifted above the pan
surface, allowing a net radial displacement to be induced, and leading
to the toroidal motion.  Note that the toroidal mode is induced below
the critical acceleration of $1.2g$ reported for fluidization of
granular matter in other studies~\cite{xxx,E90,E92}.  For $\phi >
30^\circ$ and $\omega>16$~Hz, the particle bed is fluidized over most
of the pan area.  This is the range of $\omega$ in which the ``wave''
and ``sector'' modes are seen.
     
In their study of a deep bed of granular matter, Wassgren et
al.~\cite{WBH96} have shown that the bed undergoes a stage with an
arched base, and during this stage, the bed moves out-of-phase with
the container. A similar argument can be put forth for the circular
bed in the present study. The motion that is imparted to the bed
depends on the pan motion and the phase difference between the pan and
the bed. When the bed moves out-of-phase with the pan, the point of
next contact with the pan becomes a function of the pan motion.  The
instantaneous traces of the pan vertical displacement show
period-doubling when the surface waves were observed in the particle
bed.  At every second cycle, the maximum upward vertical displacement
is slightly less than that of the first cycle.  The difference in
amplitude decreases as the surface waves accelerate and the sectors
eventually appear.  The sectors may thus correspond to a standing
wave.
    
Our observations are thus consistent with the existence of well
defined states based on the frequency of the excitation.  Computer
simulations are currently underway to further elucidate the origin of
the behavior described here~\cite{BDSMP02}.

\begin{acknowledgments}
We thank Materials and Manufacturing Ontario and NSERC (Canada) for
financial support.  PHP also acknowledges support from the Canada
Research Chairs program. We also acknowledge valuable discussions with
K. Bevan., S. Fohanno, J.J. Drozd and E.B. Smith.

\end{acknowledgments}

\newpage

\end{document}